\documentclass[aps,twoside,onecolumn,showpacs,nofootinbib]{revtex4-2}

\usepackage{amssymb}
\usepackage{amsmath}
\usepackage{bm,multirow}
\usepackage{verbatim}
\usepackage{hyperref}
\usepackage{subfigure}
\renewcommand{\arraystretch}{1.3}
\usepackage[pdftex]{graphicx}

\begin{document}

\title{Quark-Meson Coupling Model in Heavy-Ion Collision Simulations}

\author{Dae Ik Kim}
\email[]{di.kim.phys@gmail.com}
\affiliation{Department of Physics, Pusan National University, Busan 46241, Korea}

\author{Chang-Hwan Lee}
\email[]{clee@pusan.ac.kr}
\affiliation{Department of Physics, Pusan National University, Busan 46241, Korea}

\author{Kyungil Kim}
\affiliation{Institute for Rare Isotope Science, Institute for Basic Science, Daejeon 34000, Korea}

\author{Youngman Kim}
\affiliation{Center for Exotic Nuclear Studies, Institute for Basic Science, Daejeon 34126, Korea}

\author{Sangyong Jeon}
\affiliation{Department of Physics, McGill University, Montreal H3A2T8, Quebec, Canada}

\author{Kazuo Tsushima}
\affiliation{Laboratório de Física Teórica e Computacional-LFTC, Programa de
P\'{o}sgradua\c{c}\~{a}o em Astrof\'{i}sica e F\'{i}sica Computacional,
Universidade Cidade de S\~{a}o Paulo, 01506-000 S\~{a}o Paulo, S\~{a}o Paulo, Brazil}

\published{Phys. Rev. C 113, 024615 (2026)}

\begin{abstract} 
The quark-meson coupling (QMC) model incorporates quark degrees of freedom into the relativistic mean-field (RMF) framework, distinguishing it from traditional quantum hadrodynamics (QHD), which treats nucleons as point-like particles. In this work, we implement the QMC model within the DaeJeon Boltzmann-Uehling-Uhlenbeck (DJBUU) transport code to investigate its applicability to intermediate-energy heavy-ion collisions. We simulate \textsuperscript{197}Au+\textsuperscript{197}Au collisions at a beam energy of 400 A MeV using both QHD and QMC and find that both approaches yield comparable results for bulk observables such as transverse and directed flow, with good agreement with experimental data. To further assess the model performance, we study pion production in neutron-rich (\textsuperscript{132}Sn+\textsuperscript{124}Sn) and less neutron-rich (\textsuperscript{108}Sn+\textsuperscript{112}Sn) systems at 270 A MeV. In contrast to the QHD case, reproducing the observed pion yields and charge ratios within the QMC framework requires a slightly reduced density-dependent suppression in the in-medium $\Delta$ production cross-section. These results demonstrate that the QMC model can be effectively integrated into transport simulations.
\end{abstract}

\maketitle

\section{Introduction} 
Understanding dense nuclear matter is strongly related to various interesting topics in nuclear physics and astrophysics, such as atomic nuclei, neutron stars and core-collapse supernovae. Various methods exist for studying nuclear matter, which can be categorized as theoretical approaches, nuclear experiments, and astronomical observations~\cite{Horowitz2014}. A major theoretical approach involves the construction of the nuclear equation of state (EOS) that determines the properties of infinite nuclear matter. The EOS is tested against constraints from nuclear experiments, such as heavy-ion collisions that produce nuclear matter~\cite{Danielewicz2002} and astronomical observations, including those of neutron stars whose outer core is considered to be uniform nuclear matter~\cite{Chatziioannou:2020pqz, Lindblom1992}.

Recently, experimental progress has been made in both fields. In heavy-ion collisions, the newly constructed rare-isotope beam facilities, such as the Radioactive Isotope Beam Factory (RIBF) in Japan, Facility for Rare Isotope Beams (FRIB) in the USA, and Rare isotope Accelerator complex for ON-line experiments (RAON) in Korea, have provided or are expected to provide useful information by generating nuclear matter with an exotic isospin ratio~\cite{Tshoo:2013voa,Jeong:2018qvi}. Since the first detection of gravitational waves~\cite{LIGOScientific:2016aoc}, multimessenger astronomy has ushered in a new era, and new constraints on the tidal deformability of neutron stars have been obtained~\cite{Abbott2018, LIGOScientific:2020zkf}.

The relativistic mean field (RMF) theory is a suitable theoretical framework for describing the nuclear EOS, which was introduced by Walecka~\cite{Walecka1974AMatter}. The Walecka model or quantum hadrodynamics (QHD) successfully explained the EOS of dense nuclear matter with the mean-field approximation and hadron degrees of freedom, although its first version showed excessively large incompressibility beyond 500 MeV. To address this drawback, Boguta and Bodmer~\cite{Boguta1977RelativisticSurface} adopted the nonlinear self-coupling of the scalar meson,  which resulted in a reasonable incompressibility that satisfied experimental constraints such as those originating from giant monopole resonances~\cite{Youngblood1999, Garg2018} and allowed its successful application to nuclear structure~\cite{Lalazissis:1996rd}.

Guichon~\cite{Guichon1988} proposed another approach to the RMF framework, called the quark-meson coupling (QMC) model, which is based on the quark degrees of freedom and mean field calculations within the MIT bag model. As an extension of QHD, QMC provides a more fundamental description of the residual strong force and naturally yields reasonable incompressibility without additional self-coupling of the scalar field. In addition, QMC has been applied to infinite nuclear matter, neutron stars, and finite nuclei~\cite{Saito1996,Guichon1996,Guichon2018,Stone2017,Saito1994}. Furthermore, in the QMC model, coupling constants with other baryons, such as the $\Delta$ baryon and hyperon, are naturally derived in the same manner as those of the nucleon. This feature allows the study of the possibility of $\Delta$ matter, neutron stars with hyperons, and hypernuclei, without requiring additional approaches or assumptions for the coupling constants of other baryons. 

To study the EOS obtained from RMF theory, within the context of heavy-ion collisions, we require an additional bridging framework. The transport model, which describes a non-equilibrium hadronic many-body system using a semi-classical approximation, is a suitable option. 
Covariant transport models such as RVUU~\cite{Ko:1988zz, Song:2015hua}, RBUU~\cite{Gaitanos:1998bu,Gaitanos:2001hv}, GiBUU-RMF~\cite{Buss2012Transport-theoreticalReactions}, and DJBUU~\cite{Kim:2020sjy} allow the study of dense nuclear matter produced in heavy-ion collisions within the RMF framework; however, most of them typically use QHD. Therefore, in heavy-ion collision simulations, QHD has already been comparatively studied with other models, including those based on the Skyrme interaction~\cite{TMEP:2016tup,TMEP:2017mex, TMEP:2021ljz, TMEP:2022xjg}, and applied to reproduce the results of heavy-ion collision experiments~\cite{Cassing:1999mh, Kummer:2023hvl},
whereas QMC has remained largely unexplored in this context.

To study the QMC model within the context of heavy-ion collisions, we implement the QMC model in DJBUU by replacing the originally implemented QHD mean field with the QMC one. For clarity, we refer to the original version of DJBUU using QHD as DJBUU+QHD and the modified version adopting the QMC model as DJBUU+QMC.

This study primarily aims to demonstrate the validity of the QMC model in heavy-ion collision simulations using a transport model. We benchmark the QMC model in heavy-ion collision simulations by simulating two sets of heavy-ion collisions corresponding to the FOPI~\cite{Reisdorf2012} and S$\pi$RIT~\cite{SpiRIT:2020sfn} experiments using DJBUU+QHD and DJBUU+QMC, respectively. 
In the first set, which consisted of the $^{197}$Au+$^{197}$Au collision system at a beam energy of 400 A MeV, we focus on flow observables, such as transverse and directed flows. These observables are sensitive to nuclear EOS and have been extensively studied using transport models~\cite{Danielewicz2002}.
In the second set, which consisted of $^{108}$Sn+$^{112}$Sn,  $^{124}$Sn+$^{112}$Sn and $^{132}$Sn+$^{124}$Sn systems at a beam energy of 270 A MeV, we focus on the pion multiplicities and their ratios. These are sensitive to the nuclear EOS as well; however, their replication remains challenging within current transport models.

The remainder of this paper is organized as follows. Sec.~\ref{sec:rmf} reviews the relativistic mean-field theory and compares the QHD approach with the QMC model. Sec.~\ref{sec:djbuu}  introduces the proposed transport model, DJBUU. The simulation results for the Au+Au collisions at 400~A MeV and Sn+Sn collisions at 270 A MeV, which focus on flow observables and pion production, are discussed in Sec.~\ref{sec:results}. Finally, in Sec.~\ref{sec:summary},
 we summarize our findings and discuss future perspectives for DJBUU+QMC, including its possible applications in heavy-ion collisions.

\section{Relativistic Mean Field Theory}
\label{sec:rmf}
RMF theory provides a strong framework for describing nuclear interactions through the exchange of mesons within the mean-field approximation. We introduce two RMF approaches: the QHD and QMC models.
\\

\subsection{Quantum hadrodynamics (QHD) model}
QHD treats nucleons as point-like particles with interactions mediated by isoscalar (scalar $\sigma$ and vector $\omega$) and isovector (vector $\rho$) meson fields. The corresponding Lagrangian density is expressed as follows:
\begin{widetext}
\begin{align}
\mathcal{L} &= \bar{\psi}\bigl[i\gamma_{\mu}\partial^{\mu} 
             - m_N + g_\sigma\sigma
             - g_{\omega} \gamma_\mu \omega^\mu 
             - g_{\rho} \gamma_\mu \vec{\tau}\!\cdot\!\vec{\rho}^{ \mu}
             - \frac{e}{2} \gamma_\mu (1+\tau_3) A^\mu\bigr]\psi 
              +  \frac{1}{2} \bigl(\partial_\mu \sigma \partial^\mu \sigma - m_\sigma^2 \sigma^2\bigr)
              -  U_\sigma(\sigma)
\notag\\
            &\quad - \frac{1}{4} \Omega_{\mu\nu} \Omega^{\mu\nu} 
                   + \frac{1}{2} m_{\omega}^2 \omega_{\mu} \omega^{\mu} 
                   - \frac{1}{4} \vec{R}_{\mu\nu}\cdot\vec{R}^{\mu\nu} 
                   + \frac{1}{2} m_{\rho}^2 \vec{\rho}_{ \mu}\!\cdot\!\vec{\rho}^{ \mu}
                   + \frac{1}{4} F_{\mu\nu} F^{\mu\nu}.
\label{eq:lagrangian_qhd}
\end{align}
\end{widetext}
Here, $m_{N}^{*} = m_{N} - g_{\sigma} \sigma$ is the effective mass of the nucleon, while the tensor field strengths of $\omega$, $\rho$ meson field, and electromagnetic field $A_\mu$ are given by
\begin{equation}
    \begin{array}{l}
        \Omega_{\mu\nu} = \partial_\mu\omega_\nu - \partial_\nu\omega_\mu,  \\
        \vec{R}_{\mu\nu} = \partial_\mu \vec\rho_\nu - \partial_\nu \vec\rho_\mu,\\
        F_{\mu\nu} = \partial_\mu A_\nu - \partial_\nu A_\mu.
     \end{array}
\end{equation}
The nucleon isospin operator $\vec{\tau}$ and its third-component $\tau_3$ is $+1$ for the proton and $-1$ for the neutron.

The scalar self-interaction $U_\sigma(\sigma)$ is typically assumed as
\begin{equation}
U_\sigma(\sigma) = \frac{1}{3} g_2 \sigma^3  +  \frac{1}{4} g_3 \sigma^4.
\end{equation}

The Euler-Lagrange equation for the nucleon field is the effective Dirac equation:
\begin{equation}
    i(\gamma_\mu(\partial^\mu-V^\mu)-(m_N-V_s))\psi=0,
    \label{eq:effectiveDiraEq}
\end{equation}
where $V^\mu = g_\omega\omega^\mu+g_\rho\vec\tau\cdot\vec\rho^\mu$ and $V_s = g_\sigma \sigma$. In nuclear matter at low temperatures, nucleons that follow Eq.~(\ref{eq:effectiveDiraEq}) can be treated as quasi-particles whose effective mass and effective momentum are $m^*=m-g_\sigma\sigma$ and $p^{*\mu}=p^\mu-V^\mu$, respectively.

Within the mean-field approximation, the mean fields for these mesons are replaced with their time components as follows:
\begin{align}
&\langle \sigma \rangle = \sigma, \\
&\langle \omega^0 \rangle = \omega, \\
& \langle \rho_3^0 \rangle = \rho
\end{align}

Solving the Euler-Lagrange equations for the meson fields using the mean-field approximation yields the following equations of motion for the meson mean field:
\begin{align}
\label{eq:sigma_eq_QHD}
&m_\sigma^2\sigma + g_2 \sigma^2 + g_3 \sigma^3 = g_{\sigma} \rho_s, \\
\label{eq:omega_eq_QHD}
&m_\omega^2\omega = g_{\omega} \rho_B, \\
\label{eq:rho_eq_QHD}
&m_\rho^2 \rho = g_{\rho} \rho_{B,I}.
\end{align}
Here, $\rho_s$, $\rho_B$, and $\rho_{B,I}=\rho_p-\rho_n$ are the scalar, baryon, and isovector densities, respectively.
\begin{align}
&\rho_s = \sum_{i=p,n}  2 \int  \frac{d^3 k}{(2\pi)^3} ~\frac{m^*}{\sqrt{k^2+{m^*}^2}} n_i(k), \\
&\rho_B = \sum_{i=p,n}  2 \int \frac{d^3k}{(2\pi)^3} ~ n_i(k), \\
&\rho_{B,I} = \sum_{i=p,n}  2 \int \frac{d^3k}{(2\pi)^3} ~ \tau_3 ~n_i(k),
\end{align}
where $n_i(k)$ is the Fermi-Dirac distribution function for nuclear matter. For cold nuclear matter, $n_i(k)$ and $k_F$ are the Heaviside step function
$\Theta(k_F-k)$ and Fermi momentum, respectively. For heavy-ion collision simulations using DJBUU, $n(k)$ is replaced with phase-space distribution $f(\vec{x},\vec{p})$ in Sec.~\ref{sec:djbuu}. 

In summary, the effective mass with a scalar potential is 
\begin{equation}
    m_B^*=m_B-g_\sigma\sigma,
\end{equation}
where $\sigma$ is obtained from the self-consistent calculation of Eq.~(\ref{eq:sigma_eq_QHD}).
In this study, we assume that $g_\sigma$ was independent of the baryon species (for nucleons and $\Delta$-resonances). The vector potential is given by
\begin{equation}
      V^0 = V^0_\omega +V^0_\rho = g_\omega\omega+g_\rho\rho\tau_3,
\end{equation}
which is obtained from Eq.~(\ref{eq:omega_eq_QHD}) and Eq.~(\ref{eq:rho_eq_QHD}).

For the DJBUU simulations, we used the coupling constant from parameter set 1 in Ref.~\cite{Liu:2001iz} which was fitted to reproduce nuclear matter properties such as the saturation density $\rho_0 = 0.16~ \mathrm{fm}^{-3}$, binding energy $E/A = -16 ~\mathrm{MeV}$, incompressibility $K_0 = 240 ~\mathrm{MeV}$, effective mass $m^* = 0.75 ~m_N$, and symmetry energy $S=30.5 ~ \mathrm{MeV}$ at $\rho_0$. 
The corresponding coupling constants are listed in Table~\ref{tab:constrants}.

\subsection{Quark-meson coupling (QMC) model}
\label{sec:qmc}
In contrast to QHD, the QMC model~\cite{Guichon2018,Saito1994,Saito2007} describes a baryon as a cluster of confined quarks, where the light quarks directly interact with external scalar and vector mean fields. This approach naturally introduces baryon-meson coupling through quark-level coupling based on the MIT bag model. 

The Lagrangian density with quark degrees of freedom is given by
\begin{widetext}
\begin{align}
    \mathcal{L}_q =& \bigl[\bar{\psi_q}(i\gamma_\mu\partial^\mu-(m_q-g^q_\sigma\sigma)-g^q_\omega\gamma_\mu\omega^\mu-g^q_\rho\gamma_\mu\vec{\tau^q}\cdot\vec{\rho}^\mu)\psi_q+\bar{\psi}_Q(i\gamma_\mu\partial^\mu-m_Q)\psi_Q-B\big]\Theta_{\mathrm{bag}}
\notag\\
            & +  \frac{1}{2} \bigl(\partial_\mu \sigma \partial^\mu \sigma - m_\sigma^2 \sigma^2\bigr)
                    - \frac{1}{4} \Omega_{\mu\nu} \Omega^{\mu\nu} 
                   + \frac{1}{2} m_{\omega}^2 \omega_{\mu} \omega^{\mu} 
                   - \frac{1}{4} \vec{R}_{\mu\nu}\cdot \vec{R}^{\mu\nu} 
                   + \frac{1}{2} m_{\rho}^2 \vec{\rho}_{ \mu}\!\cdot\!\vec{\rho}^{ \mu},
\label{eq:lagrangian_quark}
\end{align}
\end{widetext}

where $\psi_q$ and $m_q$ are the light quark fields $u$, $d$ and their masses, respectively; while $g^q_\sigma$, $g^q_\omega$, and $g^q_\rho$ are the quark-meson coupling constants; Quarks and their interactions are restricted in the baryon bag by Heaviside step function $\Theta_{\mathrm{bag}}$. $B$ is the bag constant. $\vec\tau^q$ is the quark isospin operator and its third-component $\tau_3^q$ is $+1$ for the $u$ quark and $-1$ for the $d$ quark. $\psi_Q$ and $m_Q$ are the heavy quark fields $s$, $c$, and $b$ and their masses, respectively. 
We assume that mesons couple only to light quarks to be consistent with the Zweig rule. 
In addition, the beam energies used in our simulations are not sufficiently high to consider heavy quarks. Therefore, we consider only light quarks at the quark level and nucleons and $\Delta$ baryons at the hadronic level.

Within the mean-field approximation, the equation of a quark field in a spherical cavity of radius $R^*_B$  is given by
\begin{equation}
    \big(i\gamma\cdot\partial-m^{*}_q-g^q_\omega\gamma_0\omega-g^q_\rho\gamma_0\tau^q_3\rho\big)\psi_q
    =0,\quad (r<R_B^*)
\end{equation}
where the effective mass of the quark is 
\begin{equation}
m_q^{*} = m_q - g^q_\sigma \sigma.
\end{equation}
The boundary condition at the surface of the bag ($r=R^*_B$) is 
\begin{equation}
\label{eq:boundary_condition}
    (1+i\gamma\cdot\hat{r})\psi_q=0.
\end{equation}
The solution for the ground state of the quark field is given by:
\begin{equation}
    \begin{aligned}
        \psi_q &= \mathcal{N}_{q}^{B *}\exp({-i\Omega^*_q t/R^*_B}) \\
        &~\times\binom{j_0\left(x_q^* r / R^*_B\right)}
    {i \beta^{B*}_q \vec{\sigma} \cdot \hat{r} j_1\left(x^*_q r / R^*_B\right)} 
    \frac{\chi}{\sqrt{4 \pi}},
    \end{aligned}
\end{equation}
where the normalization factor is:
\begin{equation}
    \mathcal{N}_{q}^{B *} =x_{q}^{*}\big(2R_B^{*3} j_0^2\left(x_{q}^*\right)\left[\Omega_{q}^*\left(\Omega_{q}^*-1\right)+m_{q}^* R_B^*/2\right] \big)^{-\frac{1}{2}}.
\end{equation}
Here, $\chi$ is the Pauli spinor, and
\begin{equation}
\Omega_q^{*} = \sqrt{ (x_q^{*})^2 + (m_q^{*} R_B^{*})^2 }, 
\end{equation}
$x_q^*$ is obtained from Eq.~(\ref{eq:boundary_condition}). The following expression is obtained:
\begin{equation}
    j_0(x^*_q)=\beta^{B*}_qj_1(x^*_q)
\end{equation}
where $j_{0}$ and $j_{1}$ are spherical Bessel functions and
\begin{equation}
    \beta^{B*}_q=\sqrt{\frac{\Omega^*_q-m_q^*R^*_B}{\Omega^*_q+m_q^*R^*_B}}.
\end{equation}

The equations of motion for the meson fields as Euler-Lagrange equations for the quark Lagrangian density are given by
\begin{align}
    (\partial_\mu\partial^\mu+m_\sigma^2)\sigma
    &=g_\sigma^q\bar{\psi_q}\psi_q,\\
    (\partial_\mu\partial^\mu+m_\omega^2)\omega^\mu &=g^q_\omega\bar{\psi_q}\gamma^\mu\psi_q,\\
    (\partial_\mu\partial^\mu+m_\rho^2)\vec{\rho}^\mu 
    &=g^q_\rho\bar{\psi_q}\gamma^\mu\tau^q_3\psi_q.
\end{align}
These equations are applied to the nuclear system $|A\rangle$ as a collection of non-overlapping bags in the external meson field~\cite{Saito2007}. Considering the Lorentz transformation from the rest frame of the bag to the nuclear rest frame, the source terms are given by
\begin{align}
    &\langle A|\bar{\psi_q}\psi_q(\vec{r})|A\rangle
    =3g_\sigma^qS(\sigma) 
    \langle A|\sum_i\frac{m^*_i}{E^*_i}\delta(\vec{r}-\vec{r_i})|A\rangle,\\
    &\langle A|\bar{\psi_q}\gamma^0\psi_q(\vec{r})|A\rangle
     =3g^q_\omega\langle A|\sum_i\delta(\vec{r}-\vec{r_i})|A\rangle,\\
     &\langle A|\bar{\psi_q}\gamma^0 \tau^q_3\psi_q(\vec{r})|A\rangle
     =g^q_\rho\langle A|\tau^q_3\sum_i\delta(\vec{r}-\vec{r_i})|A\rangle, 
\end{align}
Here, the scalar integral of the quark field corresponding to the inner structure of the baryon is
\begin{equation}
S(\sigma) = \int_{\mathrm{bag}} d^3 r \bar{\psi}_q \psi_q=\frac{\Omega^*_q/2+m^*_qR^*_B(\Omega^*_q-1)}{\Omega_{q}^*\left(\Omega_{q}^*-1\right)+m_{q}^* R_B^*/2}.
\end{equation}
Assuming that the terms containing the time and spatial derivatives are negligible, we obtain the hadronic expression for the equations of motion for the meson mean field with baryon-meson coupling constants from the quark-meson coupling constants: 
\begin{align}
    \label{eq:Meson_equations_QMC}
    m_\sigma^2\sigma&=g_\sigma C_B(\sigma)\rho_s,\\
    \label{eq:Meson_equations_QMC_omega}
    m_\omega^2\omega&=g_\omega\rho_B,\\
    \label{eq:Meson_equations_QMC_rho}
    m_\rho^2\rho &= g_\rho\rho_{B,I},
\end{align}
where $C_B=S(\sigma)/S(0)$. The quark-meson coupling constants ($g^q_\sigma$, $g^q_\omega$, and $g^q_\rho$) and baryon-meson coupling constants ($g_\sigma$, $g_\omega$, and $g_\rho$) are related as follows
\begin{equation}
    g_\sigma = 3g^q_\sigma S(0), \quad g_\omega =3g^q_\omega,\quad  g_\rho =g^q_\rho.
\end{equation}
Notably, $g_\sigma$ has different meanings in QMC and QHD. $g_\sigma C_B(\sigma)$ in Eq.~\ref{eq:Meson_equations_QMC} acts like the density-dependent coupling $g_\sigma(\sigma)$, $g_\sigma$ in QMC is $g_\sigma(0)$, so $g_\sigma$ in QHD corresponds to $g_\sigma C_B(\sigma)$, whereas $g_\omega$ and $g_\rho$ in QMC have the same meaning as those in QHD.

The effective mass of a baryon $B$ including nucleons and $\Delta$ resonances in nuclear matter arises from the bag energy in the external meson mean fields given by
\begin{equation}
m_B^{*} = 
\sum_{q} \frac{n_q \Omega_q^{*} - z_0}{R_B^{*}} 
+ \frac{4}{3} \pi (R_B^{*})^3 B,
\label{eq:bag_energy}
\end{equation}
subject to the stability condition:
\begin{equation}
\frac{d m_B^{*}}{d R_B^{*}} = 0.
\label{eq:QMC_BC}
\end{equation}
Here, $z_0$ accounts for the center-of-mass and gluon fluctuation corrections and $B$ is the bag constant. Both of them are assumed to be independent of the density. $n_q$ denotes the number of  light quarks in the baryon.

For practical calculations, $m^*_B$ is expressed as a power series expansion in $g_\sigma \sigma$, truncated at second order~\cite{Saito2007,Guichon2018,Tsushima:2022PTEP, Choi2021}:
\begin{equation}
        m_B^* = m_B-g_\sigma\sigma +\frac{a_B}{2}(g_\sigma \sigma)^2,
\end{equation}
where $a_B$ is obtained from self-consistent quark-level calculations with Eq.~(\ref{eq:bag_energy}) and depends on baryon species, because it is determined by the bare masses of baryons also. For our simulations, we consider only nucleons and $\Delta$ baryons.
Their masses are, 939 MeV and 1232 MeV, respectively. Consequently, $a_N$ = 0.181 fm for nucleons and $a_\Delta$ = 0.199 fm with a light quark mass $m_q$ of 5 MeV and bag constant $B$ = (170 MeV)$^{4}$ in the parameter set
found in Ref.~\cite{Tsushima:2022PTEP}.

Finally, we discuss how to obtain mean-field potentials. As the way to obtain scalar potential, we introduce two options. The first one is solving Eq.~(\ref{eq:Meson_equations_QMC}) self-consistently with $C_B(\sigma)$ linearly parameterized with $a_B$:
\begin{equation}
C_B(\sigma) = 1 - a_B (g_{\sigma}\sigma).
\end{equation}
The alternative way is using density-dependent parameterizations suggested by Tsushima~\cite{Tsushima:2022PTEP}.
The parameterization for $g_\sigma \sigma$ (in MeV) is given by
\begin{equation}
(g_\sigma\sigma)(x) = 
\begin{cases}
1.608  -  23.91 ~\sqrt{x}  +  350.6~ x \\
\quad - 144.3 ~x \sqrt{x}  +  19.48 ~x^2 & (x>0)\\
0 & (x=0)
\end{cases}
\end{equation}
where $x=\rho_B/\rho_0$, $\rho_0=0.15$ fm$^{-3}$. This assumption is valid when $0<x<3.0$. 
It allows us to determine $g_\sigma\sigma$ without involving scalar density, thereby reducing computational cost.

The parameterizations for vector potential (in MeV) is given by
\begin{equation}
    V^0 = V^0_\omega + V^0_\rho = b_B~x + 84.61 y \left(\frac{\tau_3}{2}\right),
\end{equation}
where $b_N=b_\Delta=125.3$, $y=(\rho_p-\rho_n)/\rho_0$. This is exactly equivalent to solving Eq.~(\ref{eq:Meson_equations_QMC_omega}) and Eq.~(\ref{eq:Meson_equations_QMC_rho}) with coupling constants in Ref.~\cite{Tsushima:2022PTEP}. 

Both options for QMC employ the same set of parameters which were calibrated at a slightly lower saturation density of $\rho_0 = 0.15~ \mathrm{fm}^{-3}$ with binding energy per nucleon of $E/A = -15.7~ \mathrm{MeV}$, incompressibility $K_0 = 280 ~\mathrm{MeV}$, and symmetry energy $S = 35 ~\mathrm{MeV}$ at $\rho_0$. 
The corresponding coupling constants are presented in Table~\ref{tab:constrants}.

\begin{table*}[t]
\centering
\renewcommand{\arraystretch}{1.5}
\setlength{\tabcolsep}{7pt} 
\begin{tabular}{cccccccclccc}
\hline 
 & $g_\sigma$ & $g_\omega$ & $g_\rho$ & $g_2$ & $g_3$ 
 & $a_N$ & $\rho_0$  &$m^*/m$& $E/A$ & $K_0$ & $S$ \\
 & & & & (fm)& 
 & (fm) & (fm$^{-3}$)  && (MeV) & (MeV) & (MeV) \\
\hline 
QHD & 8.96 & 9.24 & 3.77 & $-4.68$ & $-30.9$ & 0    & 0.16  &0.75&       $-16$&       240&       30.5\\
QMC & 8.23 & 8.15 & 4.67 & 0     & 0     & 0.181 &      0.15 &0.8&       $-15.7$&       280&       35\\
\hline
\end{tabular}
\caption{Coupling constants and nuclear matter properties for the QHD~\cite{Liu:2001iz} and QMC~\cite{Tsushima:2022PTEP} models used in the DJBUU model.}
\label{tab:constrants}
\end{table*}

\section{DaeJeon Boltzmann-Uehling-Uhlenbeck Model}
\label{sec:djbuu}
The DJBUU model is based on the relativistic Boltzmann-Uehling-Uhlenbeck (BUU) equation given by
\begin{equation}
\label{eq:relbuu}
\frac{1}{E_i^{*}}\left[p^\mu \partial_\mu^x - \Bigl(p_\mu \mathcal{F}^{\mu\nu} - m_i^*   \partial_x^\nu m_i^*\Bigr)\partial_\nu^p\right] f_i(\vec{x},\vec{p}) = \mathcal{C}_i(\vec{x},\vec{p}),
\end{equation}
where $p^\mu$, $\mathcal{F}^{\mu\nu}=\partial^\mu V^\nu-\partial^\nu V^\mu$, and $m_i^*$ denote the four-momentum, vector field tensor, and effective mass of a particle of species $i$ in the RMF approach, respectively. 
The effective energy is $E_i^*=\sqrt{\vec{p}^2+m^{*2}_i}$ and $f_i(\vec{x}, \vec{p})$ represents the phase-space density of the species $i$. When the left-hand side of Eq.~(\ref{eq:relbuu}) is equal to 0, it corresponds to the Vlasov equation. The right-hand side, $\mathcal{C}_i$, describes the collisions of the two baryon species $i$ and $j$, including elastic and inelastic collisions with Pauli blocking. If we denote the incoming particles as 1 and 2 and the outgoing particles as 3 and 4, the collision term for $12\leftrightarrow34$ is given by
\begin{equation}
\begin{aligned}
\mathcal{C}_{12\leftrightarrow34}&(\vec{x}, \vec{p})= \frac{1}{2} \int \frac{d^3 p_2}{(2 \pi)^3 2 p_2^0} \int \frac{d^3 p_3}{(2 \pi)^3 2 p_3^0} \int \frac{d^3 p_4}{(2 \pi)^3 2 p_4^0} \\
 \times & \left|\mathcal{M}_{12\leftrightarrow34}\right|^2(2 \pi)^4 \delta\left(p_1+p_2-p_3-p_4\right) \\
 \times &\left\{f_3\left(\vec{x}, \vec{p}_3\right) f_4\left(\vec{x}, \vec{p}_4\right)\left[1-f_1\left(\vec{x}, \vec{p}_1\right)\right]\left[1-f_2\left(\vec{x}, \vec{p}_2\right)\right]\right. \\
& \left.-f_1\left(\vec{x}, \vec{p}_1\right) f_2\left(\vec{x}, \vec{p}_2\right)\left[1-f_3\left(\vec{x}, \vec{p}_3\right)\right]\left[1-f_4\left(\vec{x}, \vec{p}_4\right)\right]\right\}.
\end{aligned}
\label{eq:collision}
\end{equation}
Here, $\mathcal{M}_{12\leftrightarrow34}$ is an element of the scattering matrix for 1+2$\leftrightarrow$3+4 scattering.

To solve Eq.~(\ref{eq:relbuu}), DJBUU employs the test-particle method~\cite{Wong1982}, where the distribution function $f(\vec{x},\vec{p})$ is represented by the sum of the shape functions of the test particles as follows:
\begin{equation}
\label{eq:tpdist}
f(\vec{x}, \vec{p}) = \frac{(2\pi)^3}{N_\mathrm{TP}}
\sum_{j=1}^{A N_\mathrm{TP}} g_x\bigl(\vec{x} - \vec{x}_j\bigr)   g_p\bigl(\vec{p} - \vec{p}_j\bigr).
\end{equation}
$A$ and $N_\mathrm{TP}$ represent the mass number of the system and the number of test particles per nucleon, respectively, while $g_x$ and $g_p$ are the shape functions defined in the coordinate and momentum spaces, respectively. If the shape function is a delta function, an excessively large number of $N_{TP}$ is required to obtain a smooth distribution. However, this approach is computationally expensive, and $N_{TP}$ is reduced by adopting a shape function with a width. In DJBUU, these functions have the following polynomial profile:
\begin{equation}
\label{eq:polyprofile}
g(\vec{u}) = \mathcal{N}_{2,3}  \bigl[1 - (|\vec{u}| / a)^2\bigr]^3
\quad \text{for } 0 < |\vec{u}| / a < 1,
\end{equation}
with $a = 4.2 ~\mathrm{fm}$, which corresponds to a width of approximately $1.4  ~\mathrm{fm}$ if a Gaussian profile is used. 

By setting $\mathcal{C}(\vec{x}, \vec{p})=0$ in the BUU equations using the test particle method, we obtain the following equation of motion for the test particles:
\begin{equation}
\frac{d \vec{x}}{d t}  =\frac{\vec{p}}{E^*},\quad
\frac{d \vec{p}}{d t}  =-\nabla V^0-\frac{m^* \nabla m^*}{E^*} .
\end{equation}
The solution of these equations yields the contribution of the mean field to the time evolution of the phase-space distribution. This process is known as propagation.

The contributions of $\mathcal{C}_i(\vec{x}, \vec{p})$ to Eq.~(\ref{eq:relbuu}) are numerically realized as hard collisions between test particles. First, for all pairs of test particles, we assess whether those in a pair are scattered. If the test particles 1 and 2 are assessed, a collision is detected when the pair satisfies the closest approach criterion, also known as Bertsch’s prescription:
\begin{equation}
    b_\mathrm{12} < \sqrt{\hat\sigma_{12}/\pi}, \quad \hat\sigma_{12}=\sigma_{12}/N_{TP}.
\end{equation}
Here, $b_\mathrm{12}$ is the transverse distance between particles 1 and 2, and $\sigma_{12}$ is the total cross-section of the elastic and inelastic collision channels between them, which is scaled as $\hat\sigma_{12}$ because of the test particle method. In this study, we use a constant cross-section of 40 mb for elastic collisions, a representative value commonly used in comparative studies~\cite{TMEP:2016tup, TMEP:2017mex, TMEP:2019yci,  Xu2024TMEP_270}. Second, a random number is generated and scattering to outgoing particles 3 and 4 is considered to have occurred if it is smaller than Pauli blocking probability $1-(1-f_3)(1-f_4)$.  We refer to this process as baryon-baryon collision.

In addition, we discuss $\Delta$ production from nucleon-nucleon collisions.
$\Delta$ production is important for understanding pions in transport simulations, because they are primarily produced through the $\Delta$-resonance channel. 
In this study, we employed the parameterization of the $N N \to N \Delta$ cross-section suggested in Ref.~\cite{Bertsch:1988ik}.
\begin{equation}
\label{eq:sigma_free}
\sigma_{N N \to N \Delta}(\sqrt{s}) =
\frac{20  (\sqrt{s} - 2.015)^2}{0.015 + (\sqrt{s} - 2.015)^2} \mathrm{mb},
\end{equation}
where $\sqrt{s}$ is above $m_{\mathrm{thr}}$, and $m_{\mathrm{thr}}$ of 2.015 GeV corresponds to $2m_N + m_\pi$.
The cross-section is further modified in the medium by a density factor corresponding to the effect, which depends on the density and isospin ratio of the nuclear medium. 
Our cross-section including in-medium modification is given by
\begin{equation}
\begin{aligned}
\label{eq:in-medium_modification}
&\sigma_{N N \to N \Delta}^*(\sqrt{s}, \rho_B, N,Z) 
\\&=
\sigma_{N N \to N \Delta}(\sqrt{s}) \times
\exp\big(-C\frac{\rho_B}{\rho_0}\big)  
\bigg(\frac{N}{Z}\bigg)^{x^{\pm,0}},
\end{aligned}
\end{equation}
where $\rho_B$ is the baryon density at the collision center, while $N$ and $Z$ are the number of neutrons and protons in the total system, respectively, with our approximation of $N/Z=\rho_n/\rho_p$ of the local collision center. 
$C$ is a constant parameter
while $x^+$, $x^0$, and $x^-$ are the isospin-dependent parameters for channels related to $\Delta^{++}$, $\Delta^{+,0}$, and $\Delta^{-}$, respectively. These have been introduced as methods to reproduce pion observables, such as the double-pion ratio measured in a S$\pi$RIT experiment~\cite{SpiRIT:2020sfn}. In particular, we used parameter set 4 of $(C,x^+,x^0,x^-)=(2.5,  0.0,  0.5,  2.0)$ from Ref.~\cite{Kim:2022sbj}.

 In summary, propagation and baryon-baryon collision are essential for transport simulations. A description of these factors for all test particles at each time step describes the complete time evolution of the dynamics of a heavy-ion collision,  including the initial, compressed, and final stages. This  enables the study of the relationship between the nuclear EOS and observables, such as collective flow and pion yield, from the compressed and final stages, respectively.

As the DJBUU code is inherently modular, the nuclear mean-field potential can be switched from the original QHD interaction to the newly implemented QMC interaction without major structural changes. We modified only the modules that computed the scalar-meson mean field and associated effective masses $m^{*}$ of the nucleons and $\Delta$ baryons, as well as those that evaluated the derivative terms in the equations of motion for the test particles, such that the baryonic force depended on the particle species. All other modules, including those for initializing the projectile and target nuclei, and for handling the nucleon-nucleon collision term, remained unchanged. For convenience, we refer to the original version with a QHD mean field as DJBUU+QHD and the variant that adopts the QMC mean field as DJBUU+QMC.

\section{Results and Discussions}
\label{sec:results}

We perform transport simulations using DJBUU+QMC and DJBUU+QHD for two sets of heavy-ion collisions. In the first set, we study \textsuperscript{197}Au+\textsuperscript{197}Au collisions at a beam energy of 400 A MeV, similar to the conditions investigated in the FOPI experiment. In the second set, we focus on $\pi$ multiplicities and ratios by comparing $\pi$ production in \textsuperscript{132}Sn+\textsuperscript{124}Sn and \textsuperscript{108}Sn+\textsuperscript{112}Sn collisions, which are inspired by S$\pi$RIT experiments.

As mentioned in Sec.~\ref{sec:qmc}, we implement two options to obtain the mean field with DJBUU+QMC: (i) solving the meson equations self-consistently through iterative calculation, and (ii) using density-dependent parameterizations fitted to the baryon density. For convenience, we refer to the former as QMC$_\mathrm{iter.}$ and the latter as QMC$_\mathrm{param.}$

\subsection{Au+Au collisions at 400 A MeV}
\label{sec:qhd}
We present the results of the simulations of \textsuperscript{197}Au+\textsuperscript{197}Au collisions at $E_{\mathrm{beam}}$ of 400 A MeV with an impact parameter $b$ = 4.7 fm corresponding to $b_0$ = 0.35, to compare with the results of \textsuperscript{197}Au+\textsuperscript{197}Au corresponding to the  $0.25<b_0< 0.45$ data of the FOPI experiment~\cite{Reisdorf2012}. The reduced impact parameter was defined as $b_0 = b/b_{\mathrm{max}}$, where $b_{\mathrm{max}} = 1.15 \times (A_P^{1/3}+A_T^{1/3})$. The number of runs and test particles per nucleon are 30 and 100, respectively, which resulted in a total of 3,000 events.

\begin{figure}[ht] 
\centering
\includegraphics[width=300pt]{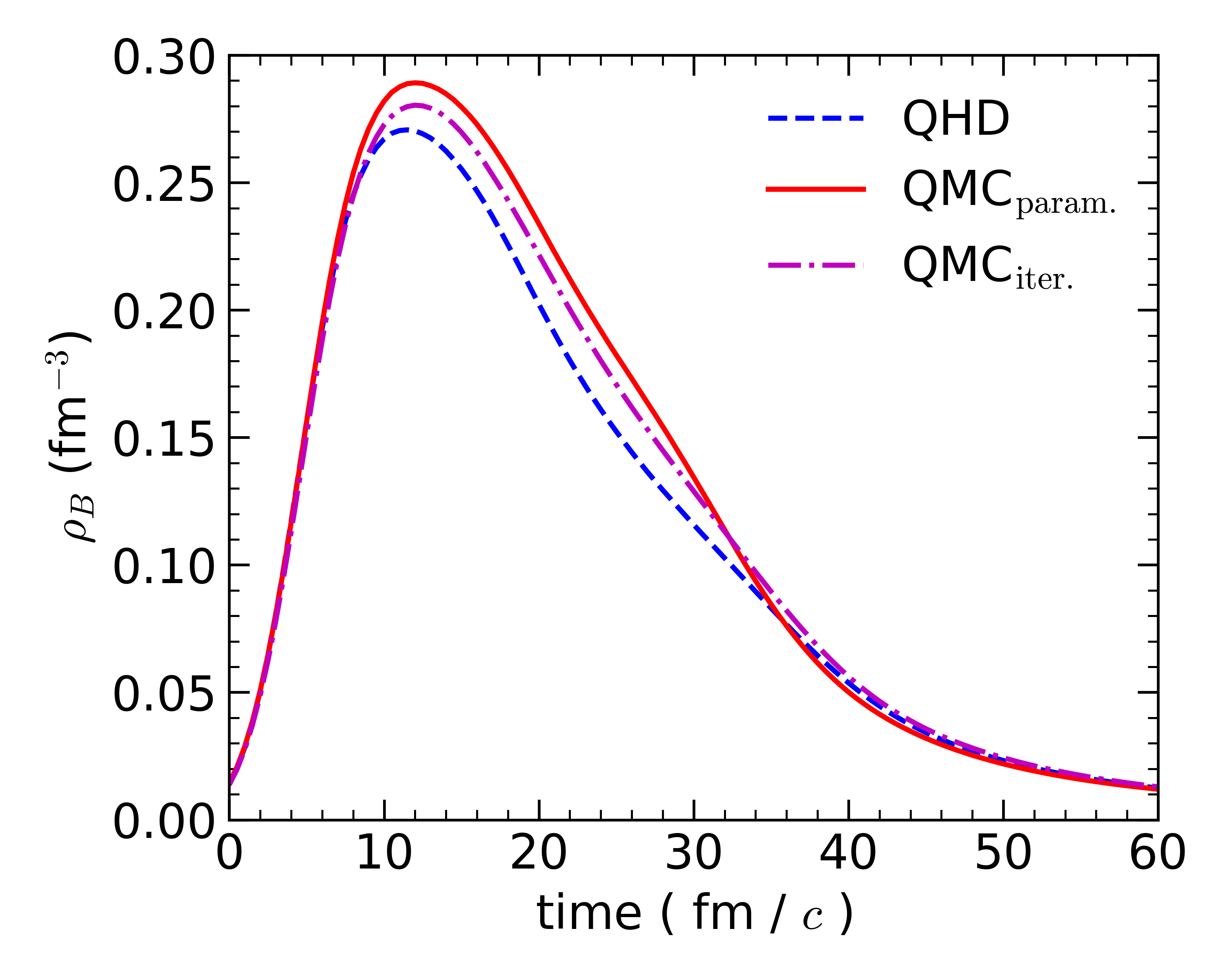}
\caption{Time evolution of the central baryon density in \textsuperscript{197}Au+\textsuperscript{197}Au collision with a beam energy of 400 A MeV.}
\label{fig:centraldensity}
\end{figure}

Figure~\ref{fig:centraldensity} shows the time evolution of the baryon density at the origin of the center-of-mass frame of the entire collision system. 
Although the central baryon density is not directly measurable in experiments, it provides valuable theoretical insights into the compression of nuclear matter.
The maximum value of the baryon density corresponds to the density of the most compressed nuclear matter produced during heavy-ion collisions. 
If some observables, such as the collective flow, were sensitive to EOS, a study of the maximum density would be meaningful from the perspective of linking the EOS to observables.  
All results exhibit similar values of the maximum density almost simultaneously at approximately 11 fm/$c$; however, the results with QMC$_\mathrm{iter.}$ and QMC$_\mathrm{param.}$ are approximately 3 and 6\% higher than those with QHD, respectively.

Given that the QMC model has a larger incompressibility ($K_0$ = 280 MeV) than QHD ($K_0$ = 240 MeV), the result appears to be somewhat counterintuitive when only $K_0$ is considered. However, many other factors besides $K_0$ can influence the maximum density. For example, a recent study~\cite{Long2024} showed that while a larger $K_0$ predicts a higher maximum density, cases with different values of $L$ can exhibit a larger maximum density even for a smaller $K_0$ because the effect of $L$ can dominate. In our case, the symmetry energy, $\rho_0$, and effective mass differ between QHD and QMC, whereas in Ref.~\cite{Long2024} the symmetry energy was varied while $\rho_0$ and the effective mass were kept fixed. These differences make a direct analysis difficult. Nevertheless, we note that the symmetry energy in the QMC model is also stiffer than in QHD, which suggests that other factors may be at play. One possible factor is the Dirac effective mass. A previous study~\cite{Choi2021-apj} investigated the relationship between the neutron star mass and radius and the nuclear matter properties such as $K_0$, symmetry energy, and the effective mass ratio $m^*/m$, defined as the nucleon effective mass at saturation density divided by its bare mass. Their results showed that a larger Dirac mass leads to a softer equation of state, and that a difference of 0.05 in $m^*/m$ has a stronger effect than a 40 MeV difference in $K_0$. If a similar trend holds in heavy-ion collision simulations, the larger effective mass in QMC may induce a stronger softening effect than in QHD. This could explain why QMC produces a larger maximum central density than QHD, even though it has a stiffer $K_0$ and symmetry energy. This interpretation, however, needs to be tested by a systematically controlled heavy-ion collision study, which is beyond the scope of the present work.

The first observable compared is the transverse flow, which is defined as $\langle p_x/A \rangle$, that is, the mean $x$-component of the momentum divided by the mass number $A$ of the emitted particles. 
In the DJBUU simulations, the momenta $p_x, p_y, p_z$ of all the protons and neutrons are tracked. The $z$-axis points along the beam direction; the $x$-axis lies along the impact parameter vector; while the $ y $-axis is perpendicular to the reaction plane. 
However, no dedicated channel for light-cluster formation (e.g., deuterons) was implemented. 
Therefore, we approximate $\langle p_x/A \rangle$ as the average transverse momentum per nucleon of the emitted particles, including light clusters, by computing the mean transverse momentum $\langle p_x \rangle$ over all the protons and neutrons.

The transverse flow is typically considered as a function of the rapidity or transverse momentum $p_t=\sqrt{p_x^2+p_y^2}$. Here, we chose reduced rapidity following Ref.~\cite{Reisdorf2012}. 
The rapidity $y$ and reduced rapidity $y_{\mathrm{beam}}$ are obtained from
\begin{equation}
    y = \frac{1}{2}\ln\left(\frac{E+p_z}{E-p_z}\right),\quad y_0=y/y_{\mathrm{beam}},
\end{equation}
where $y_{\mathrm{beam}}$ is the rapidity of the beam at the center-of-mass frame:
\begin{equation}
    y_{\mathrm{beam}} =\frac{1}{2}~ \cosh^{-1}(1+E_{\mathrm{beam}}/m_N).
\end{equation}

\begin{figure}[ht] 
\centering
\includegraphics[width=300pt]{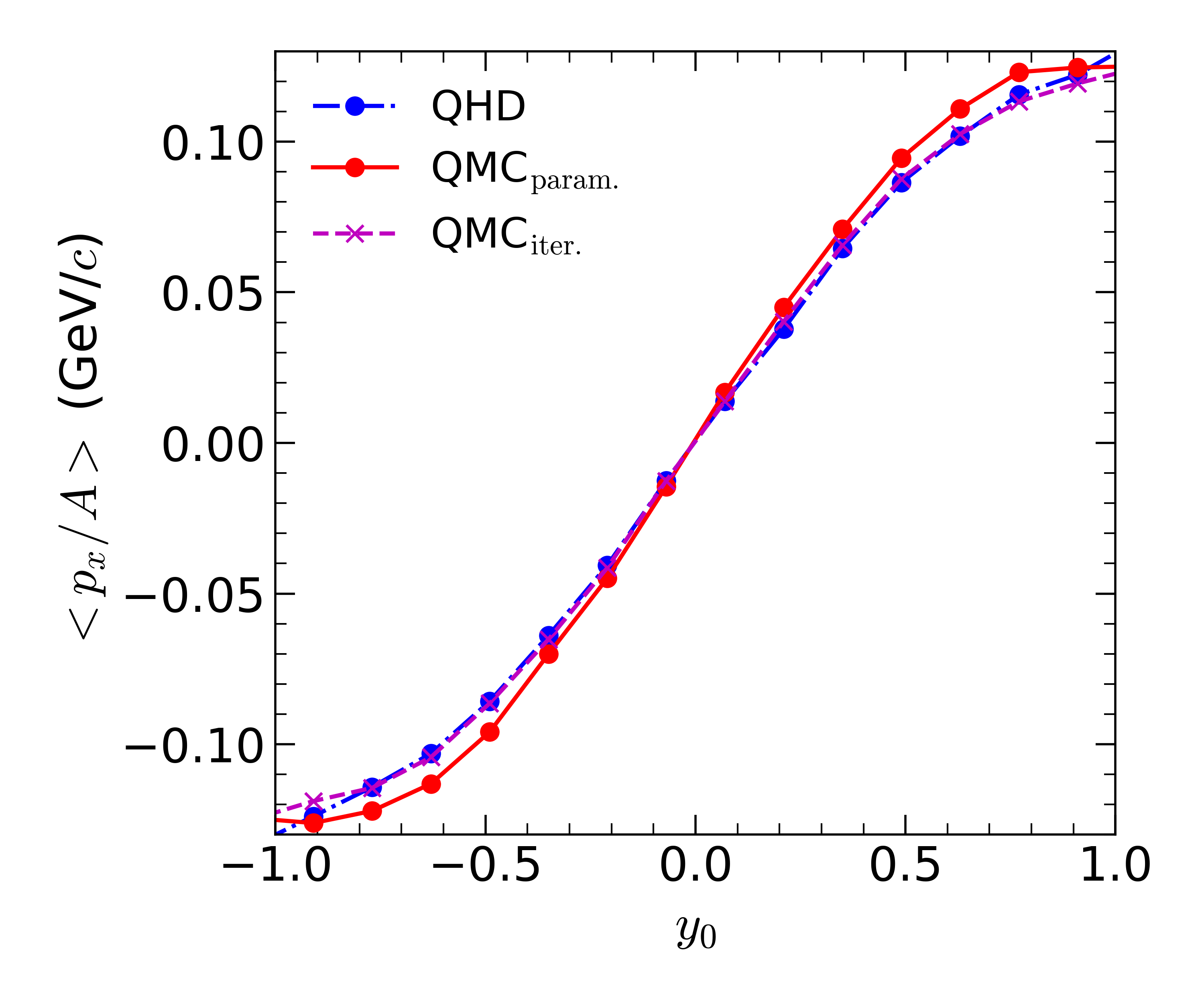}
\caption{Transverse flow $\langle p_x \rangle$ or $\langle p_x/A \rangle$ as a function of the reduced rapidity.}
\label{Fig:transverseFlow}
\end{figure}

Figure~\ref{Fig:transverseFlow} shows the transverse flow $\langle p_x \rangle$ as a function of the reduced rapidity $y_0$.
While QMC$_\mathrm{iter.}$ and QHD follow similar trends, QMC$_\mathrm{param.}$ exhibits a slightly larger deviation relative to the others, as indicated by the slope parameter $\frac{d\langle p_x/A \rangle}{dy_0}|_{y_0=0}$ being slightly larger than that for QHD and QMC$_\mathrm{iter.}$.
This minor difference is expected, as the density-dependent parameterization for the scalar strength yields a slightly weaker attractive force than the iterated approach, resulting in a slightly larger flow slope.

After the results are validated by comparing DJBUU+QHD with DJBUU+QMC, we examine the directed flow of free protons and assess whether QMC reproduced experimental data in Ref.~\cite{Reisdorf2012}.
The directed flow is the first coefficient of the Fourier expansion for the azimuthal distribution of the emitting particles: 
\begin{equation}
    \frac{dN}{dy p_t dp_t d\phi} =\frac{dN}{dy p_tdp_t}[1+2v_1\cos(\phi)+2v_2\cos(2\phi)+\cdots],
\end{equation}
and is obtained by
\begin{equation}
    v_1 = \langle \frac{p_x}{p_t} \rangle = \langle \cos\phi\rangle. 
\end{equation}

Free protons are distinguished using
 the phase-space coalescence model, where the particles are $|\Delta\vec{x}|>R_0$ and $|\Delta\vec{p}| > P_0$. We selected $R_0 = 3.6\ \mathrm{fm}$ and $P_0 = 0.35\ \mathrm{GeV}/c$ at a freeze-out time of $60\ \mathrm{fm}/c$ based on the parameter ranges reported in Ref.~\cite{Li2016}. To use this method within the test particle method, we scale $R_0\rightarrow R_0/(N_{TP})^{1/3}$.

\begin{figure}[ht] 
\centering
\includegraphics[width=300pt]{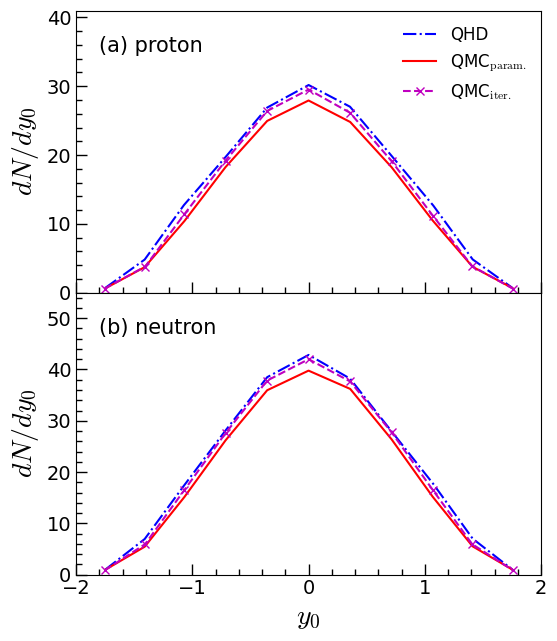}
\caption{$dN/dy_0$ of free protons and neutrons as functions of the reduced rapidity. } %
\label{fig:dndy}
\end{figure}

Figure~\ref{fig:dndy} shows the rapidity distribution of free protons and neutrons. 
Although the trends depending on rapidity are similar, the results obtained using QMC show fewer particles across the entire rapidity range for both protons and neutrons compared to those obtained using QHD. The deviation is more pronounced when comparing QMC$_\mathrm{param.}$ to QHD than when comparing QMC$_\mathrm{iter.}$.


\begin{figure}[ht]
\centering
\includegraphics[width=300pt]{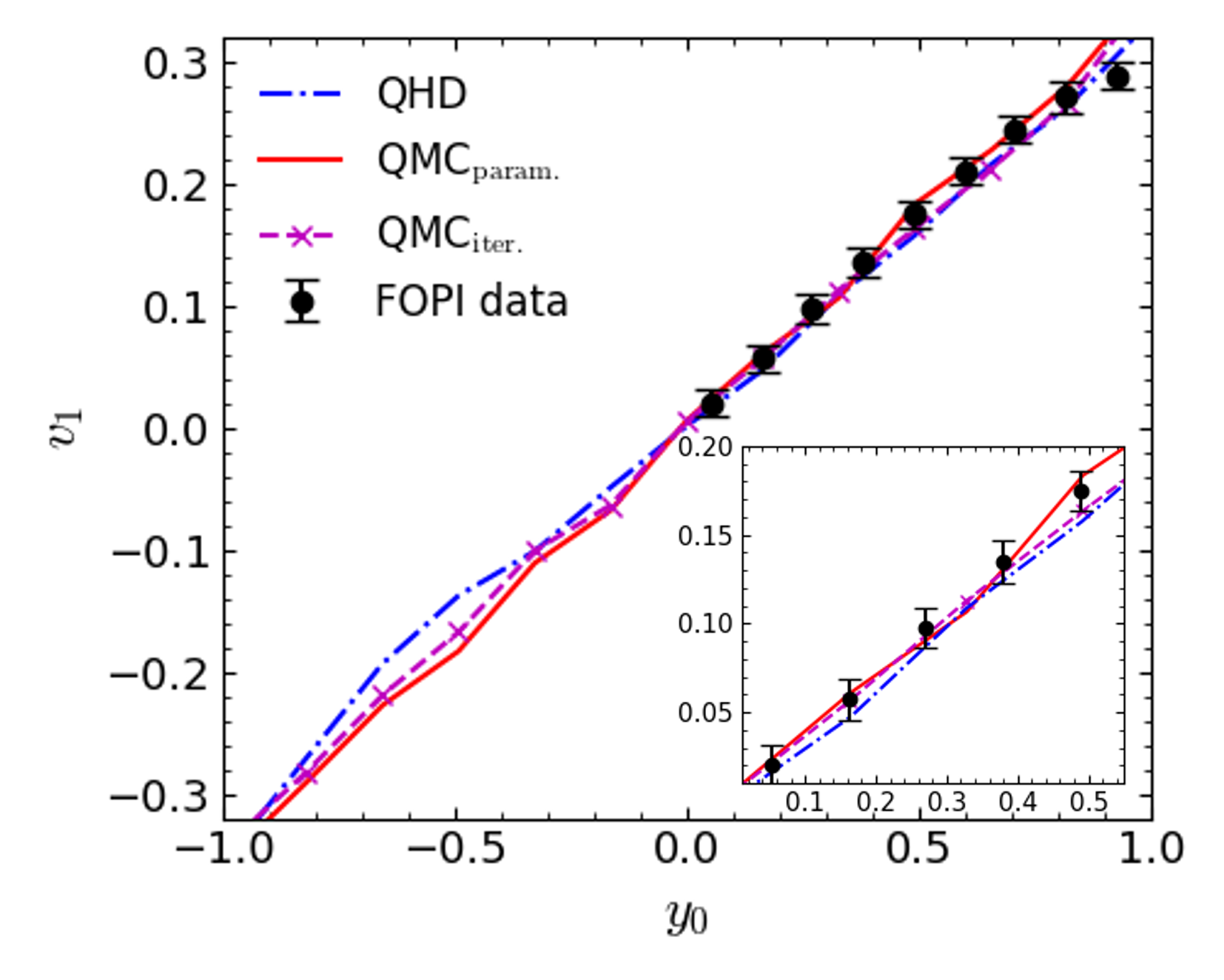}
\caption{Directed flow $v_1$ of protons in \textsuperscript{197}Au+\textsuperscript{197}Au collisions at $E_{\mathrm{beam}}$ of 400 A MeV and $b_0=0.35$. The black dots with error bars indicate the experimental data in Ref.~\cite{Reisdorf2012}.}
\label{fig:v1_auau}
\end{figure}

Figure~\ref{fig:v1_auau} shows the directed flow $v_1$ of free protons. 
The black dots with error bars indicate the experimental data of $^{197}$Au+$^{197}$Au at 400 A MeV with cuts of $0.25<b_0< 0.45$ and $u_{t0}>0.4$. Here, $u_{t0}=u_t/u_p$, $u_t$ is the transverse component of the four-velocity of the emitting particles, and $u_p$ is the beam direction component of the four-velocity of the projectile.  
The blue dot-dashed, red solid lines and magenta dashed crosses represent the results of the same system with QHD, QMC$_\mathrm{param.}$ and QMC$_\mathrm{iter.}$, respectively.
The results obtained with QMC$_\mathrm{param.}$ and QMC$_\mathrm{iter.}$ exhibit a very slightly higher value of $v_1$ in the region $y_0>0$ than that with QHD.  
All models successfully reproduce the experimental data with the same criterion for identifying free protons.
DJBUU+QMC exhibits comparable behavior in the model‐to‐model comparison and accurately describes the measured observables, thus validating its applicability to heavy-ion collisions.

\subsection{Sn+Sn collisions and isospin effects}

\begin{figure*}[ht] 
\centering
\includegraphics[width=450pt]{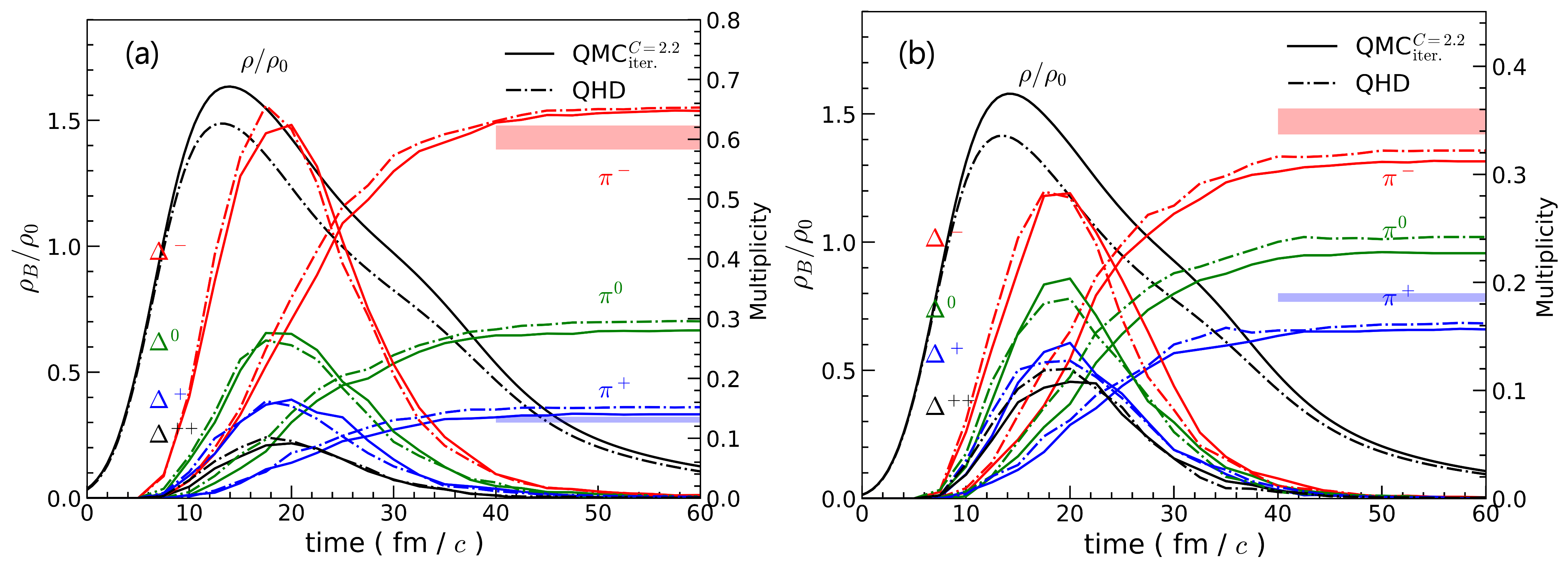}
\caption{Time evolution of the central baryon density and the multiplicities of $\Delta$ isobars and pion triplets in \textsuperscript{132}Sn+\textsuperscript{124}Sn with $N/Z = 1.56$ (a) and \textsuperscript{108}Sn+\textsuperscript{112}Sn with $N/Z = 1.2$ (b). The solid and dot-dashed lines represent the results obtained using QMC$^{C=2.2}_\mathrm{iter.}$ and QHD, respectively. For clarity, the results from QMC$_\mathrm{param.}$ and QMC$^{C=2.2}_\mathrm{param.}$ are omitted in the figure. Note that the saturation density $\rho_0$ is set to 0.16 fm$^{-3}$ for QHD and 0.15 fm$^{-3}$ for QMC.} 
\label{fig:SpiPion} 
\end{figure*}

\begin{table*}[ht]
\centering
\setlength{\tabcolsep}{12pt}
\resizebox{\textwidth}{!}{%
  \begin{tabular}{cccccccc}
    \hline
    \multirow{2}{*}{} 
      & \multicolumn{3}{c}{(a) ${}^{132}\mathrm{Sn}+{}^{124}\mathrm{Sn}\,(N/Z=1.56)$}
      & \multicolumn{3}{c}{(b) ${}^{108}\mathrm{Sn}+{}^{112}\mathrm{Sn}\,(N/Z=1.2)$}
      & \multirow{2}{*}{DR} \\
      & $Y(\pi^-)$ & $Y(\pi^+)$ & SR 
      & $Y(\pi^-)$ & $Y(\pi^+)$ & SR 
      & \\ 
    \hline
    w/ QHD
      & 0.655(18) & 0.153(10) & 4.78(45) 
      & 0.322(8) & 0.163(7) & 2.04(9)
      & 2.34(10) \\
    w/ QMC$_{\mathrm{param.}}$
      & 0.573(15) & 0.118(6) & 5.14(31) 
      & 0.259(11) & 0.136(10) & 2.11(18)
      & 2.44(10) \\
    w/ QMC$_{\mathrm{param.}}^{C=2.2}$
      & 0.686(14) & 0.152(10) & 4.99(40)
      & 0.328(13) & 0.156(9) & 2.23(15)
      & 2.23(11) \\
    w/ QMC$_{\mathrm{iter.}}^{C=2.2}$
      & 0.652(15) & 0.139(7) & 5.05(43)
    & 0.313(12) & 0.157(7) & 2.10(15)
    & 2.40(11) \\
    Exp. 
      & 0.603(20) & 0.131(5) & 4.60(11)
      & 0.349(12) & 0.186(8) & 1.89(4)
      & 2.44(10) \\
    \hline
  \end{tabular}%
}
\caption{Pion yields ($\pi^-$ and $\pi^+$), single ratios (SR = $Y(\pi^-)/Y(\pi^+)$), and double ratios (DR) for neutron-rich ($N/Z = 1.56$) and less neutron-rich ($N/Z = 1.2$) Sn+Sn collisions. Results from DJBUU calculations with QHD, QMC$_\mathrm{param.}$, QMC$^{C=2.2}_\mathrm{param.}$, and QMC$^{C=2.2}_\mathrm{iter.}$ are compared with experimental data from S$\pi$RIT~\cite{SpiRIT:2020sfn}. Both QHD and QMC$_\mathrm{param.}$ employ the same in-medium modification parameters, including a suppression factor with $C = 2.5$, whereas QMC$^{C=2.2}_\mathrm{param.}$ and QMC$^{C=2.2}_\mathrm{iter.}$ use the same modification form but with a reduced value of $C = 2.2$.}
\label{tab:results_sn}
\end{table*}

Pion observables such as pion multiplicities and ratios are considered to be strongly related to the nuclear matter produced because most pions are generated at the compressed stage~\cite{Xu2019TransportCollisions}.
To study pion production with different isospin ratios, that is, the number of neutrons divided by one of the protons in the system, Sn+Sn experiments using Sn isotope beams were performed with an S$\pi$RIT detector at the RIBF. 
Several transport models had predicted pion observables, including the double-pion ratio, which is the $\pi^-/\pi^+$ from the neutron-rich system of \textsuperscript{132}Sn+\textsuperscript{124}Sn ($N/Z = 1.56$) divided by that from the less neutron-rich system of \textsuperscript{108}Sn+\textsuperscript{112}Sn ($N/Z = 1.2$). It is expressed as
\begin{equation}
\label{eq:double_ratio}
\mathrm{DR} = 
\frac{[Y(\pi^-)/Y(\pi^+)]_{\mathrm{132+124}}}
     {[Y(\pi^-)/Y(\pi^+)]_{\mathrm{108+112}}}.
\end{equation}
However, these models showed limited predictive power prior to the availability of experimental data~\cite{SpiRIT:2020sfn}. 

To improve agreement with experimental data, various transport models have incorporated additional physics inputs, such as momentum-dependent potentials and in-medium potentials for pions and $\Delta$ resonances~\cite{Xu2024TMEP_270}. In a previous study using DJBUU, an in-medium modification of $\Delta$ production with density and isospin dependent factors described in Sec.~\ref{sec:djbuu} was introduced in DJBUU+QHD to reproduce the double ratio (DR). In the present study, we examine how DJBUU+QMC describes pion observables using the same in-medium cross-section as in DJBUU+QHD, and discuss whether both models, which consistently reproduce the $v_1$ observable, yield similar implications for pion production.

Motivated by this, we perform simulations of \textsuperscript{132}Sn+\textsuperscript{124}Sn and \textsuperscript{108}Sn+\textsuperscript{112}Sn collisions at a beam energy of 270 A MeV and an impact parameter of $b = 3.0$ fm. The number of runs and test particles per nucleon are set to 20 and 100, respectively, resulting in 2,000 events.
We examine four cases: QHD, $\mathrm{QMC}_{\mathrm{param.}}$, $\mathrm{QMC}_{\mathrm{param.}}^{C=2.2}$, and QMC$_{\mathrm{iter.}}^{C=2.2}$. In $\mathrm{QMC}_{\mathrm{param.}}^{C=2.2}$, and $\mathrm{QMC}_{\mathrm{iter.}}^{C=2.2}$, we adopt $C$ = 2.2 in the in-medium $NN\rightarrow N\Delta$ cross-section in Eq.~(\ref{eq:in-medium_modification}), while $\mathrm{QMC}_{\mathrm{param.}}$ employs the same $C$ = 2.5 as in QHD.

Figure~\ref{fig:SpiPion}
 shows the time evolution of the baryon density at the collision center and the number of pions and $\Delta$ baryons. 
The density evolution obtained with $\mathrm{QMC}_{\mathrm{param.}}$ and
$\mathrm{QMC}_{\mathrm{param.}}^{C=2.2}$ is almost identical to that of
$\mathrm{QMC}_{\mathrm{iter.}}^{C=2.2}$, with only a marginally larger maximum central density.
For clarity, we therefore omit $\mathrm{QMC}_{\mathrm{param.}}$ and
$\mathrm{QMC}_{\mathrm{param.}}^{C=2.2}$ from Fig.~\ref{fig:SpiPion}.
Overall, all QMC calculations yield a larger maximum central density than QHD, similar to what is observed in Au+Au collisions. 
The number of $\Delta$ baryons peaks at 20 fm/$c$ and subsequently decreases as they decay into pions. The number of pions increases until about 40 fm/$c$ and then saturates; therefore, we take the pion number at 60 fm/$c$ as the pion yields. In Fig.~\ref{fig:SpiPion}, we show the pion results only for QHD and $\mathrm{QMC}_{\mathrm{iter.}}^{C=2.2}$, while Table~\ref{tab:results_sn} lists the pion yields for all four cases, including $\mathrm{QMC}_{\mathrm{param.}}$  and $\mathrm{QMC}_{\mathrm{param.}}^{C=2.2}$.

 First, we compare results obtained with $\mathrm{QMC}_{\mathrm{param.}}$ and QHD, which use the same coefficient $C$ = 2.5, tuned for DJBUU+QHD. As shown in Table~\ref{tab:results_sn}, the results with QMC$_\mathrm{param.}$ yield fewer $\Delta$ baryons and pions than those with QHD, despite exhibiting a higher central density
 which is typically expected to induce more collisions and enhance the production of $\Delta$ baryons and pions. 
This unexpected outcome is attributed to the strong density suppression factor applied to the in-medium cross-section in Eq.~(\ref{eq:in-medium_modification}), which takes the form $e^{-C\rho_B/\rho_0}$ tuned for DJBUU+QHD to reproduce the experimental data and leads to excessive suppression at high densities, thereby reducing the production of $\Delta$ baryons and pions in the QMC case.

These results indicate that the in-medium modification tuned for DJBUU+QHD is not directly applicable to other models such as QMC. Nevertheless, despite the reduced pion yields, QMC reproduces the double pion ratio reasonably well. Therefore, we refit the density suppression factor for QMC by adjusting only the density-dependent part, while keeping the isospin-dependent part unchanged. 
Moreover, the effective-mass scaling approach~\cite{ems-Persram2002, ems-Li2005, ems-Cozma2021}, another prescription for in-medium modification, suggests that the cross-section scales with the effective mass as $\sigma^*/\sigma_\mathrm{free} \approx (m^*/m)^2$. Since the QMC model gives $m^*/m \approx$ 0.8, larger than 0.75 in QHD, the effective-mass scaling would predict a weaker suppression of the in-medium cross-section for QMC. These considerations imply that the coefficient $C$ may need to be readjusted to account for the model dependence associated with the effective mass. With the refitted value of $C = 2.2$, the pion yields obtained with QMC$^{C=2.2}_\mathrm{param.}$ and QMC$^{C=2.2}_\mathrm{iter.}$ become more consistent with the experimental data for both systems.

Table~\ref{tab:results_sn} summarizes the calculated pion multiplicities, single ratios (SR = $Y(\pi^-)/Y(\pi^+)$), and double ratios (DR) from DJBUU calculations with QHD, QMC$_\mathrm{param.}$, QMC$^{C=2.2}_\mathrm{param.}$, and QMC$^{C=2.2}_\mathrm{iter.}$, alongside the experimental data. As mentioned earlier, the $\pi^-$ and $\pi^+$ multiplicities are lower in the case of QMC$_\mathrm{param.}$, although the DR is well reproduced by both QHD and QMC$_\mathrm{param.}$.

QMC$^{C=2.2}_\mathrm{param.}$ yields a slightly higher $\pi^-$ multiplicity than QHD, exceeding the experimental value by more than 10\% in the neutron-rich system. Its DR also shows a small deviation compared to the result before reducing the suppression factor. In contrast, QMC$^{C=2.2}_\mathrm{iter.}$ provides better agreement with QHD and the experimental data, both in terms of pion multiplicities and DR.

Nevertheless, as in the case of QHD, QMC$^{C=2.2}_\mathrm{iter.}$ tend to slightly overestimate the pion yields in the neutron-rich system and underestimate them in the less neutron-rich  system. This suggests that further improvement in the modeling of pion production is required for both QHD- and QMC-based approaches to achieve more accurate comparisons with experimental data.

\section{Summary and Outlook}
\label{sec:summary}

In this study, we apply the quark-meson coupling (QMC) model, a quark-level relativistic mean-field approach, within the framework of the DJBUU covariant transport model. In the QMC calculation, the Dirac effective mass is obtained by including a quadratic term in $g_\sigma\sigma$, with a baryon species dependent coefficient $a_B$, which is absent in the QHD model and is derived from quark-level calculations. 
In the QMC implementation, we test two schemes: one based on self-consistent iterative solutions of the meson fields including the $a_B$ coefficients, and another using density-dependent parameterizations of the scalar potential fitted to $\rho_B$.

To benchmark validity and performance of the QMC model in heavy-ion collision simulations, we compared the results from DJBUU+QMC against those obtained using the nonlinear QHD model. We describe \textsuperscript{197}Au+\textsuperscript{197}Au collisions at beam energy of 400 A MeV. 
Simulations revealed that DJBUU with QMC produced a larger central density at the origin of the center-of-mass frame than that with QHD. A comparison of the transverse and directed flows $v_1$ obtained from QHD, QMC$_{\mathrm{iter.}}$ and QMC$_{\mathrm{param.}}$ indicates that the data from QMC match those from QHD and experimental data, whereas QMC$_{\mathrm{param.}}$ shows a slightly larger transverse flow, due to a weakened attractive force.
Subsequently, we perform \textsuperscript{132}Sn+\textsuperscript{124}Sn and \textsuperscript{108}Sn+\textsuperscript{112}Sn collisions at 270 A MeV, focusing on the central density and observable pions.
Due to the large value of the parameter for the suppression of $\Delta$ production depending on density, the results for QMC indicate that a larger central density reduces the production of $\Delta$ baryons and pions compared to those for QHD.
Although QMC reduces pion multiplicities, 
it reproduces the experimentally measured double ratio as well as QHD. Therefore, modifying the density-dependent coefficient $C$ in the in-medium modification for the inelastic cross-section from 2.5 to 2.2 enables DJBUU+QMC to reproduce pion multiplicities and their ratios more consistently with the experimental data. In particular, QMC$^{C=2.2}_\mathrm{iter.}$ shows slightly better agreement than QMC$^{C=2.2}_\mathrm{param.}$, indicating the importance of a self-consistent treatment of the meson fields.

By comparing the models and experimental data, we successfully demonstrate the feasibility of applying a nuclear mean-field potential based on the QMC model to intermediate-energy heavy-ion collisions. This study serves as a foundation for exploring the role of quark degrees of freedom in both low- and intermediate-energy heavy-ion collisions, where such effects have traditionally been considered negligible. 
Building on this work, future studies will employ the QMC model to investigate additional quark-level phenomena in dense nuclear matter, such as the in-medium baryon magnetic moments~\cite{Tsushima:2022PTEP} and the effects of short-range quark-quark correlations~\cite{Saito:2000gi}, thereby further extending the predictive power and relevance of the QMC framework in nuclear physics.
\\

\section{Acknowledgments}

D.I.K and C.-H.L were supported by the National Research Foundation of Korea (NRF) grant funded by the Korean government (No. RS-2023-NR076639).
D.I.K was supported by the Hyundai Motor Chung Mong-Koo Foundation and by the 2023 BK21 FOUR Graduate School Innovation
Support funded by the Pusan National University (PNU-Fellowship Program).
K.K. and Y.K. were supported in part by the Institute for Basic Science (2013M7A1A1075764, IBS-I001-01, IBS-R031-D1).
S.J.~acknowledges the support of the Natural Sciences and Engineering Research Council of Canada (NSERC) [SAPIN-2024-00026].
K.T.~was supported by Conselho Nacional de
Desenvolvimento Cient\'{i}fico e Tecnol\'ogico (CNPq, Brazil), Processes No. 304199/2022-2, and
FAPESP Process No.~2023/07313-6, and his work was also part of the projects, Instituto Nacional de
Ci\^{e}ncia e Tecnologia - Nuclear Physics and Applications (INCT-FNA), Brazil, Process
No.~464898/2014-5.

\bibliographystyle{apsrev4-2}
\bibliography{biblio}

\end{document}